\documentclass[lettersize,journal]{IEEEtran}

\usepackage{amsmath,amssymb,amsfonts}
\usepackage{algorithm}
\usepackage{algpseudocode}
\usepackage{graphicx}
\usepackage{booktabs}
\usepackage{tabularx}
\usepackage[table,dvipsnames]{xcolor} 
\usepackage[caption=false,font=normalsize,labelfont=sf,textfont=sf]{subfig}
\usepackage{stfloats}
\usepackage{balance}
\usepackage{verbatim}
\usepackage[hyphens]{url}
\usepackage{textcomp}
\usepackage{orcidlink}
\usepackage{ragged2e}  
\usepackage{tikz}
\usetikzlibrary{shapes, arrows, positioning, fit, calc}

\newcolumntype{L}[1]{>{\RaggedRight\arraybackslash}p{#1}}
\newcolumntype{Y}{>{\centering\arraybackslash}X}
\newcolumntype{R}[1]{>{\RaggedRight\arraybackslash}p{#1}}

\usepackage{listings}

\lstdefinelanguage{Bithoven}{
  morekeywords={pragma, version, target, return, if, else, verify, older, after, checksig,
                bool, signature, number, string, true, false, pubkey, max, min, negate, abs,
                sha256, ripemd160, len},
  sensitive=false,
  morecomment=[l]{//},
  morestring=[b]",
  keywordstyle=\color{blue}\bfseries,
  commentstyle=\color{gray!70!black},
  stringstyle=\color{purple},
  basicstyle=\ttfamily\footnotesize,
  breakatwhitespace=false,
  breaklines=true,
  captionpos=b,
  keepspaces=true,
  showspaces=false,
  showstringspaces=false,
  showtabs=false,
  tabsize=2
}
\lstdefinelanguage{Rust}{
  morekeywords={
    as, break, const, continue, crate, else, enum, extern, false, fn, for, if, impl,
    in, let, loop, match, mod, move, mut, pub, ref, return, self, Self, static, struct,
    super, trait, true, type, unsafe, use, where, while, async, await, dyn
  },
  sensitive=true,
  morecomment=[l]{//},
  morecomment=[s]{/*}{*/},
  morestring=[b]{"},
  keywordstyle=\color{blue}\bfseries,
  commentstyle=\color{gray!70!black}\itshape,
  stringstyle=\color{purple},
  basicstyle=\ttfamily\footnotesize,
  breaklines=true,
  showstringspaces=false,
  tabsize=2
}

\newcommand{\compilererror}[1]{%
  \par\noindent
  \colorbox{red!10}{%
    \parbox{0.98\linewidth}{%
      \texttt{\textcolor{red!80!black}{\textbf{Error:}} #1}%
    }%
  }%
}

\newtheorem{defn}{Definition}
\newtheorem{theorem}{Theorem}

\def\BibTeX{{\rm B\kern-.05em{\sc i\kern-.025em b}\kern-.08em
    T\kern-.1667em\lower.7ex\hbox{E}\kern-.125emX}}

\begin{document}

\title{Bithoven: Formal Safety for Expressive Bitcoin Smart Contracts}

\author{Hyunhum Cho\orcidlink{0000-0001-9256-5794}, Ik Rae Jeong\orcidlink{0000-0002-4120-2165}
\thanks{
Manuscript received;
(Corresponding author: Ik Rae Jeong.)}
\thanks{
Hyunhum Cho, Ik Rae Jeong are with the School
of Cybersecurity, Korea University, Seoul 02841, South Korea (email:
hyunhum@korea.ac.kr; irjeong@korea.ac.kr).}}

\maketitle

\begin{abstract}
The rigorous security model of Bitcoin’s UTXO architecture often comes at the cost of developer usability, forcing a reliance on manual stack manipulation that leads to critical financial vulnerabilities like signature malleability, unspendable states and unconstrained execution paths. Industry standards such as Miniscript provide necessary abstractions for policy verification but do not model the full imperative logic required for complex contracts, leaving gaps in state management and resource liveness. This paper introduces Bithoven, a high-level language designed to bridge the gap between expressiveness and formal safety. By integrating a strict type checker and a resource liveness analyzer with a semantic control-flow analyzer, Bithoven eliminates major categories of consensus and logic defects defined in our fault model prior to deployment. Our results indicate that this safety comes at modest cost: Bithoven compiles to Bitcoin Script with efficiency comparable to hand-optimized code, demonstrating that type-safe, developer-friendly abstractions are viable even within the strict byte-size constraints of the Bitcoin blockchain.
\end{abstract}


\section{Introduction}

\IEEEPARstart{S}{mart} contracts---programs that execute on a blockchain---have redefined decentralized applications. Yet, this paradigm remains sharply divided. On one hand, the account-based model, exemplified by Ethereum, offers Turing-complete expressiveness but at the cost of catastrophic bugs, re-entrancy attacks, and compiler flaws \cite{sol-bug, byte-bug, hevm}. This has spawned a complex ecosystem of post-hoc security tools \cite{dafny}.

On the other hand, the Unspent Transaction Output (UTXO) model, pioneered by Bitcoin \cite{satoshi}, prioritizes security and parallelism. Its native smart contract language, Bitcoin Script, is a simple, non-Turing-complete, stack-based language. This deliberate limitation minimizes the attack surface, but the simplicity is a double-edged sword. Writing raw Bitcoin Script is notoriously difficult, error-prone, and challenging to analyze \cite{bitcoinorg-tx}. Empirical analyses of the Bitcoin blockchain have shown that this complexity leads directly to demonstrable fund loss, uncovering thousands of defective scripts with critical vulnerabilities such as ``unbound-txid'' (anyone-can-spend) or ``useless-sig'' (mangled logic) \cite{bshunter}.

This leaves a significant gap in the design space: how can developers build sophisticated, high-assurance contracts on Bitcoin without inheriting the dangers of Turing-completeness or the esoteric pitfalls of raw script?

The research community has proposed several solutions. Formalisms like BitML are theoretical and often impractical, requiring either complex predicates or consensus changes \cite{bitml, rich-utxo}. Low-level languages like Simplicity aim to replace the VM entirely but are not designed for human-readable development \cite{simplicity}. The current state-of-the-art, Miniscript, is a monumental step forward for analyzability and composition \cite{miniscript, bip379}. However, Miniscript is primarily a highly-analyzable \textit{assembly language} and compiler target, not a developer-centric, imperative source language.

We argue for a pragmatic middle ground---one that prioritizes safety and verifiability without sacrificing developer productivity. Developers need a high-level, expressive language that is intuitive to write, yet compiles down to formally safe and verifiable Bitcoin Script.

In this paper, we present Bithoven, a new, formally specified smart contract language for Bitcoin. Bithoven provides high-level, human-readable abstractions for common cryptographic and time-locking operations, resembling familiar imperative languages. The core of our contribution is not just the language, but its correct-by-construction static analyzer. By integrating a rigorous type system \cite{BitcoinWikiScript}, a liveness and scope analyzer, a recursive security checker, and a semantic control-flow analyzer, the Bithoven compiler provides \textit{a priori} safety guarantees, eliminating entire classes of vulnerabilities before deployment.

Our contributions are as follows:
\begin{itemize}
    \item We introduce a formally specified, type-safe, high-level imperative language designed exclusively for native Bitcoin Script, including its complete syntax, operational semantics, and type system (Section \ref{sec:model} \& \ref{sec:bithoven-type-system}).
    
    \item We describe the design and implementation of Bithoven's multi-stage static analysis pipeline, which is uniquely capable of detecting not only type-system and consensus-level defects, but also complex liveness errors (e.g., signature reuse) that Miniscript does not prevent (Section \ref{sec:compiler}).
    
    \item We evaluate Bithoven's security guarantees, demonstrating its ability to prevent a wide range of documented Bitcoin Script vulnerabilities (such as all relevant classes from \cite{bshunter}) at compile time (Section \ref{sec:evaluation}).
    
    \item We provide an open-source compiler with Web IDE\cite{BithovenRepo}, implemented in Rust using established libraries\cite{lalrpop, rust-bitcoin}, and present a comparative analysis with Miniscript. We demonstrate that Bithoven provides superior expressiveness for complex logic while maintaining structural isomorphism with optimized policy languages, incurring low on-chain overhead (Section \ref{sec:evaluation-expressiveness}).
\end{itemize}

\section{Related Work}

We position Bithoven within the broader research landscape. This section is structured around three key areas: (1) fundamental blockchain programming paradigms, (2) the landscape of languages and compilers for Bitcoin, and (3) alternative application and scalability solutions for Bitcoin.

\subsection{Blockchain Programming Paradigms}

The design of a smart contract language is fundamentally constrained by the underlying ledger model.

\textbf{Account-Based Model:} 
The Ethereum model is the most prominent, treating contracts as stateful objects in a global state \cite{e-utxo}. 
This Turing-complete model provides high expressiveness, but its complexity is a major source of security vulnerabilities. 
This has led to a rich field of research in post-hoc verification, including bytecode analysis \cite{byte-bug}, compiler bug detection \cite{sol-bug}, formal semantics \cite{dafny}, and symbolic execution frameworks \cite{hevm}.

\textbf{UTXO-Based Model:} 
The UTXO model, introduced by Bitcoin \cite{satoshi}, is a stateless, parallel model where transactions consume existing outputs and create new ones\cite{BitcoinWikiScript}. 
This design is inherently more scalable and simpler to analyze, but it presents challenges for stateful computation and data storage \cite{bzip}. 
The complexity of Bitcoin's transaction structure itself has triggered various research in formal modeling to enable robust analysis \cite{bitml, formal-btc-tx}. 
Bithoven is designed as a UTXO-native language, embracing its constraints to provide stronger safety. Recent proposals such as BitVM2 and Taproot Covenants \cite{bitvm2, taproot-covenants} extend the boundaries of expressiveness within the UTXO model, underscoring the timeliness of high-level, safe design efforts such as Bithoven.

\subsection{Languages and Compilers for Bitcoin}

The primary challenge in Bitcoin smart contracts is bridging the gap between the developer ecosystem and the low-level, error-prone Bitcoin Script\cite{miniscript}.
Empirical analyses have shown that script-level complexity contributes directly to fund loss and audit difficulty \cite{bshunter}. Indeed, the Bitcoin developer community itself recognizes the significant risks of raw Script, warning that its complexity can lead to demonstrable fund 
loss \cite{bitcoinorg-tx}. This has motivated the development of several high-level solutions, most notably the Miniscript policy language \cite{miniscript, bip379}.

\textbf{Bitcoin Script Analysis:} 
The need for better tooling is motivated by analyses of the Bitcoin blockchain, which have uncovered thousands of defective scripts with vulnerabilities like ``unbound-txid'' or ``useless-sig'' \cite{bshunter}. These defects have led to a demonstrable loss of funds.  Bithoven's static analyzer is designed to prevent these specific, known defect classes by construction. Moreover, Bithoven's analyzer extends beyond these semantic 
flaws to provide a comprehensive safety model, preventing fundamental classes of type-system errors (e.g., arithmetic on a string), consensus-rule violations (e.g., malformed or off-curve public keys), and control-flow defects (e.g., non-terminal execution paths).

\textbf{Formal Languages and Calculi:} A significant body of academic work has explored formalisms for Bitcoin contracts. BitML, for example, is a process calculus for specifying contract logic, focusing on symbolic
and a computational model at the transaction level \cite{bitml}. Other work has explored the secure compilation of stateful contract logic into the UTXO model, often using a series of transactions to manage state \cite{rich-utxo}. While these works provide a strong theoretical foundation, their abstraction away from script-level specifics make them less adaptive to the rapidly evolving Bitcoin protocol (e.g., Taproot upgrade) or even require consensus extensions, reducing their immediate practicality. Bithoven builds upon insights from these theoretical models but with a pragmatic focus on the script itself, ensuring safety while simultaneously offering superior expressiveness and adaptability. This adaptability is demonstrated through features like the \texttt{pragma target} directive (for \texttt{legacy}, \texttt{segwit}, or \texttt{taproot} compilation) and a semantic model that maps directly to Bitcoin's native opcodes.

\textbf{Practical Language Proposals:} 
Several languages have been introduced to improve safety or developer experience in Bitcoin smart contracting.
\begin{itemize}
    \item \textbf{Simplicity} \cite{simplicity} is a low-level, formally-defined functional language intended to replace Bitcoin Script entirely. Its primary goal is formal provability and verification, but its combinator-based design is not applicable to intuitive, human-readable contract development.
    Bithoven's pragmatic design contrasts with Simplicity. Rather than replacing Bitcoin's consensus-critical VM, Bithoven provides high-level abstractions that compile to currently-deployed and optimized Bitcoin Script. This allows Bithoven to provide a priori safety guarantees without requiring the consensus change or Bitcoin Script extension.
    \item \textbf{Miniscript} \cite{miniscript, bip379} defines a structured, composable policy subset of Bitcoin Script. It offers strong support for automated script analysis and composition, but is best understood as a highly-analyzable assembly language and compiler target, not as a developer-centric source language. Miniscript has seen significant adoption, including its integration into the Bitcoin Core\cite{bitcoin-core}. Its specification is standardized as Bitcoin Improvement Proposal (BIP) 379 \cite{bip379}, which operates as an informational standard on top of existing consensus rules.
    \item \textbf{Descriptor} \cite{bip380}: Output Script Descriptors are a specification language used by wallets to unambiguously describe how to derive scriptPubKeys and locate relevant UTXOs. A primary benefit of descriptors is providing a unified abstraction layer that separates the wallet's high-level spending policy from the specific, versioned script implementation (e.g., P2SH, P2WPKH, or P2TR). For example, a descriptor can contain a Miniscript policy, abstracting the complex final script into a single, human-readable string. Bithoven's \texttt{pragma target} directive is directly inspired by this design, providing a similar compiler-level abstraction that ensures adaptability to future protocol upgrades.
    \item \textbf{Ivy} \cite{ivy}, \textbf{Clarity} \cite{clarity}, and \textbf{sCrypt} \cite{scrypt} bring higher-level programmability and different programming models to Bitcoin or UTXO systems. However, unlike Bithoven, these languages do not provide formal, correct-by-construction static analysis with guaranteed elimination of script-level defects prior to deployment.
\end{itemize}

\subsection{Application and Scalability on Bitcoin}

Despite the complexity and limitations of Bitcoin Script, a significant body
of research has explored methods to extend Bitcoin's capabilities.

\textbf{On-Chain Applications:} 
Researchers have long demonstrated that complex applications are possible on 
Bitcoin, including lotteries \cite{btc-lottery} and cryptographic 
commitment schemes \cite{btc-affine}. 
These applications often require complex, hand-crafted scripts---precisely 
the kind of fragile logic Bithoven aims to eliminate through structured 
abstractions.

\textbf{Scaling and Off-Chain Logic:} 
The most prominent scaling solution, the Lightning Network, moves the bulk 
of transactions off-chain \cite{lightning}. 
Other systems, like FastKitten, use trusted execution environments (TEEs) 
to run complex contracts off-chain, using Bitcoin only as a final 
settlement layer \cite{fastkitten}.

\textbf{Turing-Completeness:} 
Very recent work, such as BitVM \cite{bitvm}, has proposed complex, 
multi-transaction protocols to achieve quasi-Turing-complete computation 
on Bitcoin.

Bithoven finds a pragmatic balance: it does not aim for 
quasi-Turing-completeness, nor does it rely on external hardware 
or off-chain layers. 
Instead, Bithoven focuses on \textit{making} the most of practical, on-chain 
contracts---such as vaults, covenants, multisig schemes, and Hashed TimeLock Contracts---safe, 
expressive, and easy to write. 
This approach directly addresses the documented flaws in Bitcoin Script 
\cite{bitcoinorg-tx, bshunter}.

\section{System and Fault Model}
\label{sec:model}

We first define our system model, which follows the standard Bitcoin UTXO architecture, and then describe our threat and fault model. The latter categorizes the developer-introduced defects that Bithoven is designed to prevent.

\subsection{System Model}
\label{sec:system-model}

Our system model adopts the standard Unspent Transaction Output (UTXO) model used by Bitcoin \cite{satoshi}. 
In this model, value is locked in \textit{output scripts} (also known as \texttt{scriptPubKey}). 
To spend a UTXO, a user must broadcast a new transaction that provides a valid \textit{input script} (or \texttt{scriptSig}) for each consumed output.

The validation of each transaction input is performed by executing Bitcoin's native smart contract language, Bitcoin Script \cite{bitcoinorg-tx}. 
The virtual machine (VM) is a simple, non-Turing-complete, stack-based interpreter. 
For each input, the VM concatenates the provided \texttt{input script} with the corresponding \texttt{output script} and executes the combined script.

Execution has two possible outcomes:
\begin{itemize}
    \item \textbf{Success:} The script executes to completion, and the final value on top of the stack is \texttt{true} (or any non-zero value). The transaction is considered valid.
    \item \textbf{Failure:} The script executes to completion and the final value on the stack is \texttt{false} (or zero), or the script aborts mid-execution (e.g., due to a failed \texttt{OP\_VERIFY} or an invalid operation). The transaction is rejected.
\end{itemize}

Our model is compatible with all standard Bitcoin scripting systems, including legacy (P2SH), Segwit (P2WSH), and Taproot (P2TR).  
Because the Bithoven compiler's final output is native Bitcoin Script, it inherently preserves Bitcoin's consensus behavior and does not alter the underlying VM semantics.

\subsection{Threat and Fault Model}
\label{sec:threat-model}

We assume a benign but fallible developer aiming to write a secure \texttt{output script}---one that evaluates to \textbf{Success} only when the spender provides the intended secrets (e.g., valid signatures or preimages). 

The faults considered here exclude network-layer or consensus-level attacks (e.g., 51\% control, double-spending, transaction censorship). Instead, our focus is on logical and semantic defects in the Bitcoin Script layer introduced by developers. 

\textbf{Adversary Model:} We assume a standard network adversary capable of observing all transactions in the mempool and on the blockchain. 
The adversary can craft arbitrary \texttt{input scripts} to attempt unauthorized spending but cannot break standard cryptographic primitives (e.g., ECDSA, SHA256)\cite{bshunter}. 
The attacker’s only advantage arises from exploiting faulty \texttt{output scripts} written by developers.

Following the classification in \cite{bshunter}, we distinguish two primary defect outcomes: (1) \textit{attacker-spendable defects}, where a script unintentionally permits unauthorized spending, and (2) \textit{never-spendable defects}, where the script becomes unredeemable even by the rightful owner. We extend this taxonomy with type system, control flow, stack discipline, and consensus rule violations, which are comparably evaluated in Section VI (see Table~\ref{tab:vulnerability-comparison}).

\subsubsection{Attacker-Spendable Defects}
These are syntactically valid scripts containing semantic flaws that enable unauthorized spending.

\begin{itemize}
    \item \textbf{Semantic Security Flaws:} The script includes a logical path that can evaluate to \texttt{true} without requiring the intended cryptographic proof. This corresponds to a failure of \textit{Semantic Security} in Table~\ref{tab:vulnerability-comparison}. A representative example is the \texttt{unbound-txid} defect \cite{bshunter}, where a missing \texttt{checksig} operation on one branch creates an “anyone-can-spend” vulnerability. Similarly, the \texttt{uncertain-sig} defect occurs when a script validates a signature against an arbitrary public key provided by the spender, rather than the owner's fixed key\cite{bshunter}.
    \item \textbf{Misuse of Cryptographic Operations:} The script misuses a cryptographic primitive, such as inverting the result of a signature check (e.g., \texttt{OP\_CHECKSIG OP\_NOT}). This defect, known as \texttt{useless-sig} in \cite{bshunter}, allows an adversary to spend funds by providing an invalid signature that still evaluates to \texttt{true}.
\end{itemize}

\subsubsection{Never-Spendable Defects}
These scripts lock funds permanently, preventing even the legitimate owner from redeeming them.

\begin{itemize}
    \item \textbf{Consensus Rule Violations:} The developer introduces malformed constants that violate Bitcoin’s consensus rules, such as number literals exceeding the 32-bit limit or invalid public key encodings. These correspond to the \texttt{consensus rules} defect class from \cite{bshunter}.
    \item \textbf{Type-System and Liveness Errors:} The script performs operations on incompatible types (e.g., arithmetic on a string) or violates stack discipline (e.g., consuming a variable twice). Miniscript prevents many such issues \cite{miniscript}, but its grammar does not track stateful resource consumption, leaving it vulnerable to liveness errors such as the reuse of a consumed variable in nested logic.
    \item \textbf{Unsatisfiable Logic:} The script contains control flow that is logically impossible to satisfy (e.g., \texttt{OP\_NOT} at last execution) or a branch lacking a return path. These correspond to the \texttt{never-true} defect from \cite{bshunter} and “Control Flow” errors in Table~\ref{tab:vulnerability-comparison}.
\end{itemize}

Bithoven’s type system, stack tracer, and semantic analyzer collectively guarantee compile-time prevention of all defect classes in this model, as summarized in Table~\ref{tab:vulnerability-comparison}.

\section{Bithoven Syntax and Semantics}

This section provides the complete formal specification of the Bithoven language. The formalisms serve as a precise blueprint for a correct-by-construction compiler and static analyzer. We not only present the rules but also explain the intuition and security guarantees that each formalism provides.

\subsection{Syntax of Bithoven}

We use the following notations: $P \in \textbf{Program}$, $p \in \textbf{Pragma}$, $\Sigma \in \textbf{StackDeclaration}$, $s \in \textbf{Statement}$, $e \in \textbf{Expression}$, $f \in \textbf{SigFactor}$, $x \in \textbf{Identifier}$, $\tau \in \textbf{Type}$, $v \in \textbf{Version}$, $t \in \textbf{Target}$, $n \in \mathbb{Z}$, $b \in \{\textbf{true}, \textbf{false}\}$, and $str \in \textbf{String}$.

The abstract syntax of Bithoven is structured hierarchically, starting from a top-level program definition and progressively detailing statements, expressions, and specialized syntactic forms for cryptographic operations, as specified in Figure \ref{fig:bithoven_syntax}.

\begin{figure*}[t]
    \hrule
    \vspace{0.5ex}

    \vspace{0.5ex}
    \noindent\textbf{Program ($P$):} \text{A Bithoven program is the top-level construct, consisting of pragmas, an input stack declaration, and a script.}
    \begin{align*}
    P \quad ::= \quad & p^+ \Sigma^+ \{s^*\}
    \end{align*}

    \noindent\textbf{Pragmas ($p$):} \text{Pragmas declare metadata for the compiler, such as language version and compilation target.}
    \begin{align*}
    p \quad ::= \quad & \textbf{pragma version } v && \text{(Language Version)} \\
    | \quad & \textbf{pragma target } t && \text{(Compilation Target, e.g., Taproot)} 
    \end{align*}

    \noindent\textbf{Stack Declaration ($\Sigma$):} \text{Defines the typed variables expected as input on the stack at the beginning of execution.}
    \begin{align*}
    \Sigma \quad ::= \quad & \textbf{Stack}(x_1:\tau_1, \dots, x_n:\tau_n)
    \end{align*}
    
    \noindent\textbf{Signature Factors ($f$):}
    \begin{align*}
        f \quad ::= \quad & (e_{sig}, e_{pk}) && \text{(Single Signature Pair)} \\
        | \quad & [m, (e_{sig_1}, e_{pk_1}), \dots, (e_{sig_n}, e_{pk_n})] && \text{(Multi-Signature Structure)}
    \end{align*}
    
    \noindent\textbf{Expressions ($e$):} \text{Expressions in Bithoven evaluate to a value that is pushed onto the stack.}
    \begin{align*}
        e \quad ::= \quad & n \mid b \mid str \mid x &&  \text{$n \in \mathbb{Z}$, $b \in \{\textbf{true}, \textbf{false}\}$, $str \in \textbf{String}$, $x \in \textbf{Identifier}$} \\
        | \quad & e_1 \oplus_{math} e_2 && \oplus_{math} \in \{\textbf{+, -, max, min}\} \\
        | \quad & e_1 \oplus_{compare} e_2 && \oplus_{compare} \in \{\textbf{==, !=,\textgreater, \textgreater=, \textless, \textless=}\} \\
        | \quad & e_1 \oplus_{logical} e_2 && \oplus_{logical} \in \{ \textbf{\&\&, \textbar\textbar } \} \\
        | \quad & \ominus_{math} e && \ominus_{math} \in \{ \textbf{negate, abs, ++, -{-}}\} \\
        | \quad & \ominus_{logical} e && \ominus_{logical} \in \{ \textbf{!}\} \\
        | \quad & \ominus_{crypto} e && \ominus_{crypto} \in \{ \textbf{sha256, ripemd160}\} \\
        | \quad & \ominus_{byte} e && \ominus_{byte} \in \{ \textbf{len} \} \\
        | \quad & \textbf{checksig}(f) && \text{(Signature Check)}
    \end{align*}
    
    \noindent\textbf{Statements ($s$):} \text{Statements perform actions and control the flow of execution but do not necessarily produce a value.}
    \begin{align*}
    s \quad ::= \quad & \textbf{if } e \text{ } \{s^*\} \textbf{ else } \{s^*\} && \text{(Conditional)} \\
    | \quad & \textbf{verify } e && \text{(Assertion)} \\
    | \quad & \textbf{older } n && \text{(Relative Timelock)} \\
    | \quad & \textbf{after } n && \text{(Absolute Timelock)} \\
    | \quad & \textbf{return } e && \text{(Terminal Expression)}
    \end{align*}
    \hrule
\caption{Abstract Syntax for Bithoven}
\label{fig:bithoven_syntax}
\end{figure*}

A \textbf{Program ($P$)} is the complete compilation unit. It begins with mandatory \textbf{Pragmas ($p$)}, which are compiler directives specifying essential metadata such as the language \texttt{version} or the compilation \texttt{target} (e.g., \texttt{Legacy}, \texttt{Segwit}, or \texttt{Taproot}). This information is crucial for generating target-specific, optimized Bitcoin Script. Following the pragmas, a mandatory \textbf{Stack Declaration ($\Sigma$)} defines the typed variables ($x:\tau$) that the script expects to find on the stack at the start of execution. This declaration serves as the contract's public interface and is the foundation for static type checking and liveness analysis. To provide an intuitive, function-signature-like interface, Bithoven's stack declarations are specified in reverse order, which the compiler then maps to Bitcoin's LIFO stack. The body of the program is a block containing a sequence of statements ($s^*$) that implement the contract's logic.

\textbf{Statements ($s$)} represent actions and control flow. The language includes standard \texttt{if/else} conditionals for branching logic and a \texttt{verify} statement for runtime assertions that halt execution and fail the transaction if the condition is false. Bithoven directly supports Bitcoin's time-locking capabilities through two dedicated statements: \texttt{older} for relative timelocks (CheckSequenceVerify) and \texttt{after} for absolute timelocks (CheckLockTimeVerify), inspired by miniscript\cite{miniscript}. Finally, the \texttt{return} statement is a terminal operation that evaluates an expression to define the script's final successful state, effectively concluding the program's execution path. By mandating that only the \texttt{return} statement produces the script's final output, other statements (like \texttt{verify} or \texttt{older}) are designed not to leave intermediate values on the stack. This design greatly simplifies stack management and eliminates an entire class of stack manipulation errors common in raw Bitcoin Script.

\textbf{Expressions ($e$)} are constructs that evaluate to a value, which is then pushed onto the stack. They include literals (integers, booleans, strings) and the variables defined in the input stack declaration. The language supports a set of binary ($\oplus$) and unary ($\ominus$) operators for common operations, which are directly mapped to bitcoin opcodes, as shown in Table \ref{tab:opcode-mapping}.

\begin{table}[!t]
\centering
\caption{Mapping of Bithoven Syntax to Bitcoin Opcodes}
\label{tab:opcode-mapping}
\begin{tabular}{@{}lll@{}}
\toprule
\textbf{Category} & \textbf{Bithoven Syntax} & \textbf{Bitcoin Opcode} \\ \midrule
\textbf{Mathematics} & \texttt{+} & \texttt{OP\_ADD} \\
 & \texttt{-} & \texttt{OP\_SUB} \\
 & \texttt{++} & \texttt{OP\_1ADD} \\
 & \texttt{-{-}} & \texttt{OP\_1SUB} \\
 & \texttt{max} & \texttt{OP\_MAX} \\
 & \texttt{min} & \texttt{OP\_MIN} \\
 & \texttt{negate} & \texttt{OP\_NEGATE} \\
 & \texttt{abs} & \texttt{OP\_ABS} \\
 \addlinespace 
\textbf{Logic \& Comparison} & \texttt{\&\&} & \texttt{OP\_BOOLAND} \\
 & \texttt{||} & \texttt{OP\_BOOLOR} \\
 & \texttt{==} & \texttt{OP\_(NUM)EQUAL} \\
 & \texttt{!=} & \texttt{OP\_(NUM)NOTEQUAL} \\
 & \texttt{<}  & \texttt{OP\_LESSTHAN} \\
 & \texttt{>}  & \texttt{OP\_GREATERTHAN} \\
 & \texttt{<=} & \texttt{OP\_LESSTHANOREQUAL} \\
 & \texttt{>=} & \texttt{OP\_GREATERTHANOREQUAL} \\
 & \texttt{!} & \texttt{OP\_NOT} \\
 \addlinespace
\textbf{Cryptography} & \texttt{sha256} & \texttt{OP\_SHA256} \\
 & \texttt{ripemd160} & \texttt{OP\_RIPEMD160} \\
 \addlinespace
\textbf{Byte Operations} & \texttt{len} & \texttt{OP\_SIZE} \\ \bottomrule
\end{tabular}
\end{table}
The core of Bitcoin's authorization logic is encapsulated in the \texttt{checksig(f)} expression.

To handle the varying argument structures of signature checks cleanly, we introduce \textbf{Signature Factors ($f$)}. A factor is a syntactic construct that groups the parameters for either a simple single-signature check (a signature and a public key pair) or a complex m-of-n multi-signature check. This abstraction allows \texttt{checksig} to serve as a unified interface for different cryptographic verification schemes, enhancing readability and maintainability.

\subsection{Semantics of Bithoven}
We define the meaning of Bithoven programs using a small-step operational semantics. This formalism describes how the state of an abstract machine evolves based purely on the language constructs, abstracting away from any specific transaction context.

\subsubsection*{Semantic Domains}
The state of our abstract machine is defined by the following components:
\begin{itemize}
    \item \textbf{Values ($v$)}: The set of runtime values, including integers, booleans, strings, signatures, and public keys. $v \in \text{Value}$.
    \item \textbf{Stack ($\sigma$)}: A sequence of values, $\sigma \in \text{Value}^*$. Pushing a value $v$ onto stack $\sigma$ is denoted $v :: \sigma$.
    \item \textbf{Runtime Environment ($\rho$)}: A mapping from identifiers to their runtime values, $\rho: \text{Identifier} \to \text{Value}$. This environment is set up at the start of execution from the typed parameters in the \text{Stack Declaration}.
    \item \textbf{Configuration}: A pair $\langle K, \sigma \rangle$, where $K$ is the code to be executed (an expression or statement) and $\sigma$ is the current stack.
    \item \textbf{Terminal States}: Execution may terminate successfully with a final stack state $\sigma_f$ or in a special failure state, denoted by $\perp$.
\end{itemize}

\subsubsection*{Operational Semantics of Expressions}
The evaluation of an expression is defined by the judgment: $\rho \vdash \langle e, \sigma \rangle \rightarrow \sigma'$. This reads: ``In the runtime environment $\rho$, evaluating expression $e$ with stack $\sigma$ results in the new stack $\sigma'$."

\noindent\textbf{Literals and Variables:}
\[
\frac{}{\rho \vdash \langle v, \sigma \rangle \rightarrow v :: \sigma} \quad
\qquad
\frac{\rho(x) = v}{\rho \vdash \langle x, \sigma \rangle \rightarrow v :: \sigma} \quad
\]

\noindent\textbf{Unary and Binary Operations:}
\[
\frac{
    \rho \vdash \langle e, \sigma \rangle \rightarrow v' :: \sigma' \quad
    v = (\ominus, v')
}{
    \rho \vdash \langle \ominus e, \sigma \rangle \rightarrow v :: \sigma'
} \quad
\]
\[
\frac{
    \begin{gathered}
    \rho \vdash \langle e_1, \sigma \rangle \rightarrow v_1 :: \sigma_1 \quad
    \rho \vdash \langle e_2, \sigma_1 \rangle \rightarrow v_2 :: \sigma_2 \quad
    \\
    v = (\oplus, v_1, v_2)
    \end{gathered}
}{
    \rho \vdash \langle e_1 \oplus e_2, \sigma \rangle \rightarrow v :: \sigma_2
} \quad
\]

\noindent\textbf{Signature Check Operation:}
The \texttt{checksig} expression evaluates its arguments and pushes a boolean onto the stack. Since the actual cryptographic verification is context-dependent and thus outside the scope of these semantics \cite{bip143,bip341}, we model this by non-deterministically producing either true or false.
\[
\frac{
    \rho \vdash \langle f, \sigma \rangle \rightarrow \langle v_{args}, \sigma' \rangle \quad
    b \in \{\textbf{true}, \textbf{false}\}
}{
    \rho \vdash \langle \textbf{checksig}(f), \sigma \rangle \rightarrow b :: \sigma'
} \quad
\]

\subsubsection*{Operational Semantics of Statements}
The execution of a statement is defined by the judgment: $\rho \vdash \langle s, \sigma \rangle \rightarrow \sigma' \text{ or } \perp$. This reads: ``In environment $\rho$, executing statement $s$ with stack $\sigma$ transitions to a new stack $\sigma'$ or fails."

\noindent\textbf{Conditional and Verify Statements:}

\[
\frac{
    \rho \vdash \langle e, \sigma \rangle \rightarrow \textbf{true} :: \sigma' \quad
    \rho \vdash \langle S_{if}, \sigma' \rangle \rightarrow \sigma''
}{
    \rho \vdash \langle \textbf{if } e \text{ } \{S_{if}\} \textbf{ else } \{S_{else}\}, \sigma \rangle \rightarrow \sigma''
} \quad
\]
\[
\frac{
    \rho \vdash \langle e, \sigma \rangle \rightarrow v :: \sigma'
}{
    \rho \vdash \langle \textbf{verify } e, \sigma \rangle \rightarrow
    \begin{cases}
        \sigma' & \text{if } v = \textbf{true} \\
        \perp & \text{if } v = \textbf{false}
    \end{cases}
} \quad
\]
The rule for the \texttt{false} branch of an \texttt{if} statement is omitted for brevity but is symmetric to when it's \texttt{true}.

\noindent\textbf{Locktime Statements:}
Like \texttt{checksig}, the locktime statements \texttt{older} and \texttt{after} are assertions whose outcomes depend on the external transaction context\cite{bip65, bip112}. We model their behavior as non-deterministic choices between success (leaving the stack unchanged) and failure (halting execution).
\[
\frac{
    op_{lock} \in \{\textbf{older}, \textbf{after}\}
}{
    \rho \vdash \langle op_{lock} \ n, \sigma \rangle \rightarrow \sigma \text{ or } \perp
} \quad
\]

\noindent\textbf{Return Statement:}
This statement is terminal. It evaluates its expression and replaces the current stack with the expression's resulting stack.
\[
\frac{
    \rho \vdash \langle e, \sigma \rangle \rightarrow \sigma'
}{
    \rho \vdash \langle \textbf{return } e, \sigma \rangle \rightarrow \sigma'
} \quad
\]


\section{Bithoven Type System}\label{sec:bithoven-type-system}
A formal type system is a crucial contribution because native Bitcoin Script lacks one. In Bitcoin's stack-based VM, every item is simply a byte array. The type of a value is determined only by the opcode that interprets it (e.g. OP\_ADD treats a byte array as a number, while OP\_VERIFY treats it as a boolean)\cite{BitcoinWikiScript}. This ambiguity is a major source of developer error, leading to defects such as applying arithmetic to a string. Such type mismatches can cause script execution to fail, resulting in ``never-spendable'' defects that permanently lock funds. Bithoven’s type system is designed to eliminate this specific class of vulnerabilities at compile time, providing an a priori safety guarantee that raw Script cannot.

The set of types in Bithoven, denoted by the notation $\tau$, is derived from the primitives available in the grammar. We also include a distinct type for public key, which is semantically different from general strings or numbers. Likewise, the sig type is introduced as a symbolic type, used only in stack declarations, which is crucial for the semantic analyzer to enforce high-level security invariants such as preventing ``unbound-txid'' spending paths.
\[
\tau \quad ::= \quad \textbf{num} \mid \textbf{bool} \mid \textbf{string} \mid \textbf{sig} \mid \textbf{pubkey}
\]

\subsection{Typing Environment}
A typing environment, $\Gamma$, maps variable identifiers to their types. It is constructed from the explicit type annotations provided in the input stack declaration of a Bithoven program.
\[
\Gamma \quad ::= \quad \emptyset \mid \Gamma, x:\tau
\]
The notation $\Gamma(x)$ denotes the type of variable $x$ in the environment $\Gamma$.

\subsection{Typing Judgments}
We define two forms of typing judgments:
\begin{enumerate}
    \item $\Gamma \vdash e : \tau$ asserts that in environment $\Gamma$, the expression $e$ is well-typed and has type $\tau$.
    \item $\Gamma \vdash s \Rightarrow \textbf{ok}$ asserts that in environment $\Gamma$, the statement $s$ is well-typed.
\end{enumerate}

\subsection{Typing Rules for Expressions and Factors}

Figure \ref{fig:bithoven_type} presents the formal typing rules that define Bithoven's static type system. These rules are the core of Bithoven's compile-time safety, ensuring that every expression is type-safe before execution. They enforce strict constraints absent in raw Bitcoin Script, such as restricting mathematical operators to num types and logical operators to bool types. The system also includes specialized rules to validate the structure of cryptographic primitives, ensuring that checksig is always used with well-formed signature and public key pairs.

\begin{figure*}[t]
\hrule
\begin{align*}
\text{(T-Num)} \quad \frac{}{\Gamma \vdash n : \textbf{num}} &&
\text{(T-Bool)} \quad \frac{}{\Gamma \vdash b : \textbf{bool}} &&
\text{(T-Str)} \quad \frac{}{\Gamma \vdash str : \textbf{string}}
\end{align*}
\begin{align*}
\text{(T-Var)} \quad \frac{\Gamma(x) = \tau}{\Gamma \vdash x : \tau}
\end{align*}
\begin{align*}
\text{(T-BinaryMath)} \quad \frac{\Gamma \vdash e_1 : \tau  \quad \Gamma \vdash e_2 : \tau \quad \tau \in \{\textbf{num}\}}{\Gamma \vdash e_1 \oplus_{math} e_2 : \textbf{num}} &&
\text{(T-Logical)}  \quad \frac{\Gamma \vdash e_1 : \tau  \quad \Gamma \vdash e_2 : \tau \quad \tau \in \{ \textbf{bool}\}}{\Gamma \vdash e_1 \oplus_{logical} e_2 : \textbf{bool}} &&
\end{align*}

\begin{align*}
\text{(T-Comparison)}
\frac{
    \Gamma \vdash e_1 : \tau \quad \Gamma \vdash e_2 : \tau \quad \tau \in \{\textbf{num}, \textbf{string}, \textbf{bool}\}
}{
    \Gamma \vdash e_1 \oplus_{compare} e_2 : \textbf{bool}
} \quad
\end{align*}
\begin{align*}
\text{(T-UnaryMath)} \quad \frac{\Gamma \vdash e : \tau \quad \tau : \textbf{num}}{\Gamma \vdash \ominus_{math} e : \textbf{num}}
&&
\text{(T-UnaryByte)} \quad \frac{\Gamma \vdash e : \tau \quad \tau : \textbf{string}}{\Gamma \vdash \ominus_{byte} e : \textbf{num}}
\end{align*}
\begin{align*}
\text{(T-UnaryCrypto)} \quad \frac{\Gamma \vdash e : \tau \quad \tau \in \{\textbf{num}, \textbf{string}, \textbf{bool}, \textbf{sig}, \textbf{pubkey}\}}{\Gamma \vdash \ominus_{crypto} e : \textbf{string}}
\end{align*}

\vspace{1ex}
\begin{align*}
(\text{T-Factor-Single}) \quad \frac{
    \Gamma \vdash e_{sig} : \textbf{sig} \quad \Gamma \vdash e_{pk} : \textbf{pubkey}
}{
    \Gamma \vdash (e_{sig}, e_{pk}) \Rightarrow \textbf{factor}
} &&
(\text{T-Checksig}) \quad \frac{
    \Gamma \vdash f \Rightarrow \textbf{factor}
}{
    \Gamma \vdash \textbf{checksig}(f) : \textbf{bool}
}
\end{align*}
\begin{align*}
(\text{T-Factor-Multi}) \quad \frac{
    \Gamma \vdash m : \textbf{num} \quad \forall i \in \{1..n\}, (\Gamma \vdash e_{s_i} : \textbf{sig} \quad \Gamma \vdash e_{p_i} : \textbf{pubkey})
}{
    \Gamma \vdash [m, (e_{s_1}, e_{p_1}), \dots, (e_{s_n}, e_{p_n})] \Rightarrow \textbf{factor}
}
\end{align*}
\hrule
\caption{Typing Rules for Expressions and Factors}
\label{fig:bithoven_type}
\end{figure*}

\subsection{Typing Rules for Statements}

\noindent\textbf{If Statement:} The condition must be boolean, and both branches must be well-typed.
\[
\text{(T-If)} \quad \frac{\Gamma \vdash e : \textbf{bool} \quad \Gamma \vdash B_{if} \Rightarrow \textbf{ok} \quad \Gamma \vdash B_{else} \Rightarrow \textbf{ok}}{\Gamma \vdash \textbf{if } e \textbf{ then } B_{if} \textbf{ else } B_{else} \Rightarrow \textbf{ok}}
\]

\noindent\textbf{Verify Statement:} The expression to be verified must be boolean.
\[
\text{(T-Verify)} \quad \frac{\Gamma \vdash e : \textbf{bool}}{\Gamma \vdash \textbf{verify } e \Rightarrow \textbf{ok}}
\]

\noindent\textbf{Locktime Statements:} To formalize the typing for both \texttt{older} (CSV) and \texttt{after} (CLTV) statements in a single rule, we introduce a notation, $op_{lock}$, which can represent either operator. A key feature of Bithoven's static analysis is to prevent consensus-level defects. Therefore, the typing rule for these statements must not only validate the operation but also enforce Bitcoin's consensus constraint that the numeric literal $n$ is a valid 32-bit unsigned integer ($0 \le n < 2^{32}$)\cite{bip65, bip112}, as defined in the following rule:
\[
\frac{
    op_{lock} \in \{\textbf{older}, \textbf{after}\} \quad n \ge 0 \quad n < 2^{32}
}{
    \Gamma \vdash op_{lock} \ n \Rightarrow \textbf{ok}
} \quad (\text{T-Locktime})
\]

\noindent\textbf{Return Statement:} The inner expression must be well-typed.
\[
\text{(T-Return)} \quad \frac{\Gamma \vdash e : \tau}{\Gamma \vdash \textbf{return } e \Rightarrow \textbf{ok}}
\]

\section{Implementation and Evaluation}
\label{sec:implementation}

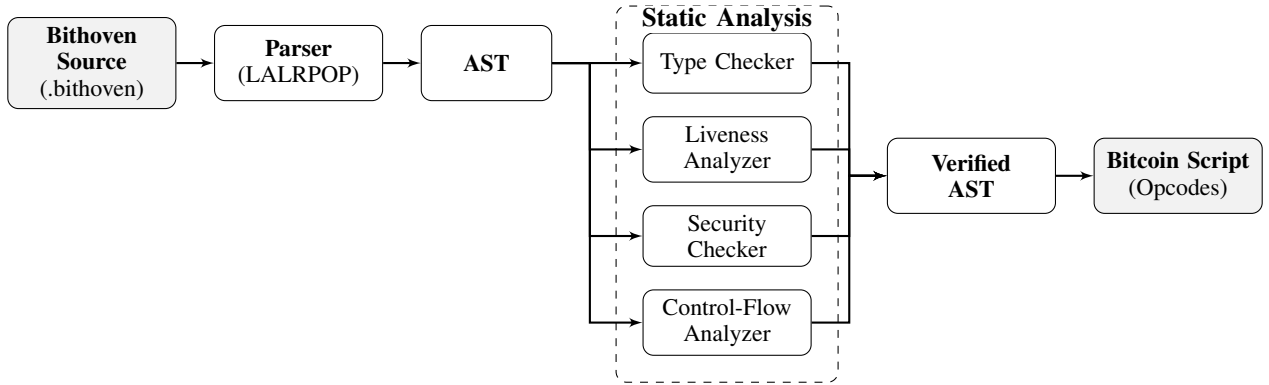
\begin{figure*}[t]
    \centering
    \begin{tikzpicture}[
        node distance=0.5cm,
        auto,
        block/.style={
            rectangle, 
            draw, 
            fill=white, 
            text width=2cm, 
            align=center, 
            rounded corners, 
            minimum height=1cm,
            font=\small
        },
        line/.style={
            draw, 
            -latex', 
            thick
        },
        dashedbox/.style={
            draw, 
            dashed, 
            inner sep=10pt, 
            label={[anchor=north, fill=white, inner sep=0.5pt]north:\textbf{Static Analysis}},
            rounded corners
        }
    ]

    
    \node (source) [block, fill=gray!10] {\textbf{Bithoven Source}\\(.bithoven)};

    \node (parser) [block, right=0.5cm of source] {\textbf{Parser}\\(LALRPOP)};

    \node (ast) [block, right=0.5cm of parser, text width=1.5cm] {\textbf{AST}};

    \node (typecheck) [block, right=1.2cm of ast, minimum height=0.8cm] {Type Checker};
    \node (liveness) [block, below=0.3cm of typecheck, minimum height=0.8cm] {Liveness Analyzer};
    \node (security) [block, below=0.3cm of liveness, minimum height=0.8cm] {Security Checker};
    \node (cfg) [block, below=0.3cm of security, minimum height=0.8cm] {Control-Flow Analyzer};

    \node (vast) [block, right=1cm of typecheck, yshift=-1.5cm, text width=2cm] {\textbf{Verified\\AST}};

    \node (output) [block, fill=gray!10, right=0.5cm of vast] {\textbf{Bitcoin Script}\\(Opcodes)};

    \node [dashedbox, fit=(typecheck) (liveness) (security) (cfg)] (analysisbox) {};

    \path [line] (source) -- (parser);
    \path [line] (parser) -- (ast);
    
    \path [line] (ast.east) -- ++(0.5,0) |- (typecheck.west);
    \path [line] (ast.east) -- ++(0.5,0) |- (liveness.west);
    \path [line] (ast.east) -- ++(0.5,0) |- (security.west);
    \path [line] (ast.east) -- ++(0.5,0) |- (cfg.west);

    \draw [line] (typecheck.east) -| ++(0.5,0) |- (vast.west);
    \draw [line] (liveness.east) -| ++(0.5,0) |- (vast.west);
    \draw [line] (security.east) -| ++(0.5,0) |- (vast.west);
    \draw [line] (cfg.east) -| ++(0.5,0) |- (vast.west);

    \path [line] (vast) -- node[above, font=\footnotesize] { } (output);

    \end{tikzpicture}
    \caption{The Bithoven Compilation Pipeline. The source code is parsed into an AST, which undergoes a multi-stage static analysis including type checking, resource liveness analysis, control flow analysis, and security verification. Only a fully verified AST is passed to the opcode generator.}
    \label{fig:architecture}
\end{figure*}

To validate our approach, we implemented a full-stack compiler and static analyzer for the Bithoven language. 
This section describes the implementation details and evaluates the system’s effectiveness in preventing the defect classes defined in our fault model (Section~\ref{sec:threat-model}).

\subsection{Implementation}
\label{sec:compiler}

The Bithoven compiler is implemented in Rust using the LALRPOP parser generator\cite{lalrpop}, which produces a formally specified parser directly from the formal grammar. 
An excerpt of this grammar is shown in Listing~\ref{list:grammar-excerpt}, illustrating the core statement and expression rules that define Bithoven’s abstract syntax tree (AST). The compilation backend to output Bitcoin Script is implemented with rust-bitcoin\cite{rust-bitcoin}, which is one of the most widely software used in bitcoin ecosystem.

\begin{lstlisting}[language=rust, caption={Bithoven LALRPOP parser grammar}, label={list:grammar-excerpt}, captionpos=b, frame=single, float=!t]
...
pub Statement: Statement = {
    <IfStatement>,
    <LocktimeStatement>,
    <VerifyStatement>,
    <ExpressionStatement>,
}

IfStatement: Statement = {
    <l:@L> "if" <c:Expression0> <b1:BlockStatement> 
    "else" <b2:BlockStatement> <r:@R> => ...
};

LocktimeStatement: Statement = 
  <l:@L> <op:LocktimeOp> <operand:UnsignedInteger> 
  <r:@R> <s:SemiColon> => ...
VerifyStatement: Statement = 
  <l:@L> "verify" <e:Expression0> <r:@R> <s:SemiColon> => ...
ExpressionStatement: Statement = 
  <l:@L> "return" <e:Expression0> <r:@R> <s:SemiColon> => ...

CheckSigExpression: Expression = {
    <l:@L> <op:CheckSigOp> <operand: SingleSigFactor> <r:@R> => ...
    <l:@L> <op:CheckSigOp> <operand: MultiSigFactor> <r:@R> => ...
}

SingleSigFactor: Factor = {
    <l:@L> <o: OpenParen> <sig:Expression4> <comma:Comma> 
    <pubkey:Expression4> <c: CloseParen> <r:@R> => ...
}
...
\end{lstlisting}

The parser outputs a typed AST, which is then passed to the static analysis pipeline composed of four core components, as illustrated in Fig. \ref{fig:architecture}:
\begin{enumerate}
    \item \textbf{Type Checker:} Enforces the type system defined in Section~\ref{sec:model}, ensuring all operations are type-correct (e.g., no arithmetic on strings) and that all cryptographic primitives are well-formed (e.g., validating public key on secp256k1 curve).
    \item \textbf{Liveness and Scope Analyzer:} Performs liveness analysis to enforce variable scoping and resource usage constraints, ensuring that resources such as signatures are consumed exactly once along any valid execution path.
    \item \textbf{Security Checker:} Performs a stateless, top-down traversal of the AST to detect local, anti-pattern vulnerabilities. This includes consensus rule violations like integer overflow and mangled operations (e.g. useless-sig vulnerability). \item \textbf{Semantic and Control-Flow Analyzer:} Traverses the program’s Control Flow Graph (CFG) to enforce complex, global invariants that cannot be found locally. This includes verifying that every execution path requires a signature (preventing \texttt{unbound-txid} defects), as well as finding control-flow defects such as a code branch lacking a \texttt{return} path. \end{enumerate}

After successful validation, the verified AST is passed to a code generator that emits optimized, target-specific Bitcoin Script (e.g., for multisig \texttt{segwit} uses \texttt{OP\_CHECKMULTISIG} while \texttt{taproot} uses \texttt{OP\_CHECKSIGADD}).

\subsection{Evaluation: Static Vulnerability Detection}
\label{sec:evaluation}

This evaluation provides the core empirical evidence for Bithoven's safety guarantees. We demonstrate that Bithoven's static analysis provides a superior safety model by comparing it on two fronts: (1) against BSHunter, a state-of-the-art \textit{post-hoc detection} tool, and (2) against Miniscript, the state-of-the-art \textit{restrictive policy language}. Table~\ref{tab:vulnerability-comparison} summarizes this analysis, showing how Bithoven provides compile-time prevention for a wider class of vulnerabilities than either alternative.

The analysis in Table~\ref{tab:vulnerability-comparison} yields two critical insights. First, it shows Bithoven's clear advantage over detection-based tools like BSHunter, which can only find existing defects; Bithoven prevents them by construction. Second, it highlights the fundamental limitations of Miniscript that Bithoven is explicitly designed to overcome:

\begin{table*}[t]
    \centering
    \caption{Comparative Analysis of Static Vulnerability Detection.}
    \label{tab:vulnerability-comparison}
    
    \begin{tabularx}{\textwidth}{@{} l X c c c X @{}}
        \toprule
        \textbf{Vulnerability Class} & \textbf{Example/Scenario} & \textbf{BSHunter} & \textbf{Miniscript} & \textbf{Bithoven} & \textbf{Key Bithoven Feature} \\
        \midrule
        
        Consensus Rules & Integer literal exceeds 32-bit limit. & $\times$ & $\checkmark$ & $\checkmark$ & Compile-time Literal Validation \\
        & Malformed public key constant. & $\checkmark$ & $\checkmark$ & $\checkmark$ & Cryptographic Type-Checking \\
        \addlinespace
        
        Type-System Safety & Applying arithmetic to a string. & $\times$ & \textit{N/A*} & $\checkmark$ & Strict Type System \\
        & Comparing a number to a string. & $\times$ & \textit{N/A*} & $\checkmark$ & Strict Type System \\
        \addlinespace
        
        Liveness \& Scope & Referencing an undefined variable. & $\times$ & $\checkmark$ & $\checkmark$ & Static Liveness \& Scope Analysis \\
        & Using a variable after it's consumed. & $\times$ & $\times$ & $\checkmark$ & Static Liveness Analysis \\
        \addlinespace
        
        Semantic Security & A code path can return \texttt{true} without a signature(\texttt{unbound-txid}). & $\checkmark$ & $\checkmark$ & $\checkmark$ & Semantic Security Analyzer \\
        & Result of a \texttt{checksig} is ignored or mangled(\texttt{useless-sig}). & $\checkmark$ & \textit{N/A*} & $\checkmark$ & Semantic Security Analyzer \\
        & \texttt{checksig} uses a public key provided by the spender(\texttt{uncertain-sig}). & $\checkmark$ & $\checkmark$ & $\checkmark$ & Semantic Security Analyzer \\
        \addlinespace
        
        Control Flow & Code exists after a \texttt{return} statement. & $\times$ & \textit{N/A*} & $\checkmark$ & Control-Flow Graph Analysis \\
        & A code branch has no \texttt{return} path. & $\times$ & \textit{N/A*} & $\checkmark$ & Control-Flow Graph Analysis \\
        
        \bottomrule
        \multicolumn{6}{l}{\textit{*N/A (Not Applicable): Feature (e.g., general arithmetic, imperative control flow) is not supported by Miniscript's policy language.}}
    \end{tabularx}
\end{table*}

\begin{enumerate} \item \textbf{State and Liveness:} Miniscript's grammar is stateless and composition-focused, which is its primary strength. However, this means it cannot track resource consumption. As the table shows, Miniscript is blind to liveness errors such as the reuse of a consumed signature. Bithoven's novel liveness analyzer \textit{is} state-aware, enforcing strict single-use constraints to eliminate this entire class of state-based errors at compile time.
\item \textbf{Securing Expressiveness:} The \textit{N/A} entries for Miniscript are not failings, but a reflection of its deliberate, restrictive design. Bithoven's core contribution is that it safely introduces these powerful, expressive imperative constructs---such as general-purpose arithmetic and \texttt{if/else} control flow. Bithoven's static analyzer is precisely what makes this new expressiveness safe, catching the very type-safety and control-flow defects that these powerful features could otherwise introduce.
\end{enumerate}

\subsection{Illustrative Examples of Defect Prevention}

We now demonstrate Bithoven’s static analyzer on representative vulnerabilities from our fault model.

\subsubsection{Semantic Security}
The most critical defect is an “anyone-can-spend” path that lacks signature authentication. 
This corresponds to the \texttt{unbound-txid} defect identified in \cite{bshunter}. 
In Listing~\ref{list:bug-nosig}, the developer mistakenly makes contract success depend only on a boolean flag.

\begin{lstlisting}[language=Bithoven, caption={An unauthenticated ``anyone-can-spend'' path.}, label={list:bug-nosig}, captionpos=b, frame=single, float=!t]
pragma bithoven version 0.0.1;
pragma bithoven target segwit;

(condition: bool)
{
    return condition == true;
}
\end{lstlisting}
\compilererror{
Error at line 4:2: NoSigRequired("At least one signature required for stack but: [StackParam { loc: Location { start: 64, end: 79, line: 4, column: 2 }, identifier: Identifier(condition), ty: Boolean }].")
}

The Bithoven compiler rejects this code because its semantic security analyzer detects a valid execution path (\texttt{condition == true}) that does not consume any input of type \texttt{signature}.

\subsubsection{State and Liveness}
A key contribution of Bithoven over Miniscript is its liveness analysis. 
A variable, especially a \texttt{signature}, is a linear resource that must be consumed exactly once. 
In Listing~\ref{list:bug-liveness}, a developer mistakenly reuses the \texttt{sig\_alice} variable in a secondary check.

\begin{lstlisting}[language=Bithoven, caption={Duplicate consumption of a signature variable.}, label={list:bug-liveness}, captionpos=b, frame=single, float=!t]
pragma bithoven version 0.0.1;
pragma bithoven target segwit;

(sig_alice: signature)
{
    verify checksig(sig_alice, <pk_alice>);
    
    // Error: sig_alice is used a second time
    return checksig(sig_alice, <pk_bob>);
}
\end{lstlisting}
\compilererror{
Error at line 9:21: VariableConsumed("Consumed variable: sig\_alice.")
}

The compiler correctly reports this as an error since the liveness analyzer tracks that \texttt{sig\_alice} was already consumed by the first \texttt{verify} statement.

\subsubsection{Consensus Rules}
Bithoven’s type checker validates all literals against Bitcoin consensus rules, preventing \texttt{impossible-key} defects \cite{bshunter}. 
In Listing~\ref{list:bug-key}, a developer provides a string that is not a valid compressed public key.

\begin{lstlisting}[language=Bithoven, caption={A malformed public key literal.}, label={list:bug-key}, captionpos=b, frame=single, float=!t]
pragma bithoven version 0.0.1;
pragma bithoven target segwit;

(sig_alice: signature)
{
    // Error: Syntactically valid 33-byte format, but point is not on the secp256k1 curve.
    return checksig(sig_alice, "0345a6b3f8eeab8e88501a9a25391318dc
    e9bf35e24c377ee82799543606bf5211");
}
\end{lstlisting}
\compilererror{
Error at line 7:32: TypeMismatch("Public key is malformed: \"0345a6b3f8eeab8e88501a9a25391318dc
e9bf35e24c377ee82799543606bf5211\"."))
}

The compiler rejects this contract, detecting that the public key violates the mathematical constraints of the secp256k1 curve required by Bitcoin consensus.
These examples collectively demonstrate that Bithoven provides \textit{a priori} safety guarantees, eliminating major classes of vulnerabilities before deployment.

\begin{table*}[!t]
    \centering
    \caption{Full Comparison: Semantics, Miniscript, and Bithoven}
    \label{tab:semantics_comparison}
    \footnotesize 
    \renewcommand{\arraystretch}{1.1} 
    
    \begin{tabularx}{\linewidth}{L{2cm} L{2.8cm} L{6cm} L{3cm} @{}}
        \toprule
        \textbf{Semantics} & \textbf{Miniscript Fragment} & \textbf{Bithoven Expression} & \textbf{Logic Overhead*} \\
        \midrule
        
        \multicolumn{4}{l}{\textit{Primitives}} \\
        False & \texttt{0} & \texttt{false} & None \\
        True & \texttt{1} & \texttt{true} & None \\
        \midrule
        
        \multicolumn{4}{l}{\textit{Key Checks}} \\
        check(key) & \texttt{pk\_k(key)} & \texttt{checksig(key)} & None \\
        check(key hash)** & \texttt{pk\_h(key)} & \texttt{verify ripemd160(sha256(k))==<H>;} \newline \texttt{return checksig(s,k');} & +1 Witness Item \\
        \midrule
        
        \multicolumn{4}{l}{\textit{Time Locks***}} \\
        nSequence $\ge$ n & \texttt{older(n)} & \texttt{older n} & +1 Opcode (\texttt{DROP}) \\
        nLockTime $\ge$ n & \texttt{after(n)} & \texttt{after n} & +1 Opcode (\texttt{DROP}) \\
        \midrule
        
        \multicolumn{4}{l}{\textit{Hash Locks**}} \\
        SHA256 & \texttt{sha256(h)} & 
        \texttt{verify len(h)==32;} \newline \texttt{return sha256(h')==<H>;} & +1 Witness Item  +1 Opcode (\texttt{DROP}) \\
        
        HASH256 & \texttt{hash256(h)} & 
        \texttt{verify len(h)==32;} \newline \texttt{return sha256(sha256(h'))==<H>;} & +1 Witness Item  +1 Opcode (\texttt{DROP}) \\
        
        RIPEMD160 & \texttt{ripemd160(h)} & 
        \texttt{verify len(h)==20;} \newline \texttt{return ripemd160(h')==<H>;} & +1 Witness Item +1 Opcode (\texttt{DROP}) \\
        
        HASH160 & \texttt{hash160(h)} & 
        \texttt{verify len(h)==20;} \newline \texttt{return ripemd160(sha256(h'))==<H>;} & +1 Witness Item +1 Opcode (\texttt{DROP}) \\
        \midrule
        
        \multicolumn{4}{l}{\textit{Combiners: AND}} \\
        Conditional & \texttt{andor(X,Y,Z)} & \texttt{if(X)\{Y\}else\{Z\}} & None \\
        Verify Seq & \texttt{and\_v(X,Y)} & \texttt{verify X; Y} & None \\
        Boolean & \texttt{and\_b(X,Y)} & \texttt{X \&\& Y} & None \\
        Wrapper & \texttt{and\_n(X,Y)} & \texttt{if(X)\{Y\}else\{return false;\}} & None \\
        \midrule
        
        \multicolumn{4}{l}{\textit{Combiners: OR}} \\
        Boolean & \texttt{or\_b(X,Z)} & \texttt{X || Z} & None \\
        Control & \texttt{or\_c(X,Z)} & \texttt{if !X \{Z\}} & None \\
        Dissatisfy & \texttt{or\_d(X,Z)} & \texttt{None} & N/A \\
        If-Else & \texttt{or\_i(X,Z)} & \texttt{if (Var) \{X\} else \{Z\}} & None \\
        \midrule
        
        \multicolumn{4}{l}{\textit{Multisig \& Threshold}} \\
        Algebraic & \texttt{thresh(k,X..)} & \texttt{(X$_1$ + X$_2$ + ..) == k} & None \\
        Multisig**** & \texttt{multi(k,key..)} & \texttt{checksig(k, [key..])} & None \\
        Tapscript**** & \texttt{multi\_a(k,key..)} & \texttt{checksig(k, [key..])} & None \\
        \midrule
        
        \multicolumn{4}{l}{\textit{Wrappers}} \\
        Alt Stack***** & \texttt{a:X} & \textit{Compiler Optimization.} & None \\
        Swap***** & \texttt{s:X} & \textit{Compiler Optimization.} & None \\
        Checksig & \texttt{c:X} & \texttt{checksig(X)} & None \\
        True & \texttt{t:X} & \texttt{X; return 1;} & None \\
        Dup & \texttt{d:X} & \texttt{None} & N/A \\
        Verify & \texttt{v:X} & \texttt{verify X} & None \\
        Wrapper & \texttt{j:X} & \texttt{if (len(Var) != 0) \{X\};} & +1 Opcode (\texttt{DROP}) \\
        Non-Zero & \texttt{n:X} & \texttt{X != 0} & None \\
        Likely & \texttt{l:X} & \texttt{if \{return 0;\} else \{X\}} & None \\
        Unlikely & \texttt{u:X} & \texttt{if (Var) \{X\} else \{return 0;\}} & None \\
        \bottomrule
        
        \multicolumn{4}{@{}p{\linewidth}@{}}{
            \footnotesize 
            \textbf{*} Logic Overhead refers to functional opcodes required by the abstraction itself. It excludes stack management instructions (e.g., \texttt{OP\_SWAP}, \texttt{OP\_TOALTSTACK}) which are generated automatically based on variable depth to ensure operand alignment and type safety. \par
            \textbf{**} Bithoven's linear resource system requires explicit witness inputs for reused variables (e.g., `h` and `h'`), adding witness overhead compared to `OP\_DUP`. \par
            \textbf{***} Bithoven time locks output an additional `OP\_DROP` (1 vByte) to enforce stack hygiene. \par
            \textbf{****} Multisig compilation output varies by target (Segwit vs Taproot). \par
            \textbf{*****} Stack wrappers are handled automatically by the compiler; resulting script size is comparable to hand-optimized Miniscript.
        }
        
    \end{tabularx}
\end{table*}

\subsection{Comparative Analysis: Semantics and On-Chain Cost}
\label{sec:comparative-analysis}

To evaluate the cost of Bithoven's high-level abstractions, we compare its imperative syntax against Miniscript policy fragments. Table \ref{tab:semantics_comparison} presents a comprehensive mapping of semantic primitives, cryptographic checks, and control flow structures, alongside a quantitative analysis of the compilation overhead.

\subsubsection{Structural Isomorphism and Zero Logic Overhead}
As evidenced by the ``Logic Overhead'' column in Table \ref{tab:semantics_comparison}, the vast majority of Bithoven constructs incur zero on-chain overhead. High-level control flow structures, such as \texttt{if/else} blocks and logical operators (\texttt{\&\&}, \texttt{||}), compile to the functionally equivalent opcode sequences as Miniscript's optimized combiners (e.g., \texttt{andor}, \texttt{or\_i}). Similarly, complex multisig and threshold checks map directly to standard \texttt{OP\_CHECKMULTISIG} or Tapscript equivalents. This demonstrates that Bithoven's imperative syntax acts as a zero-cost abstraction for the bulk of smart contract logic, preserving the efficiency of the underlying VM without imposing a ``virtual machine tax.''

\subsubsection{The Safety-Efficiency Trade-off}
The comparison highlights two specific areas where Bithoven introduces minor overhead. These are deliberate design choices where the language prioritizes formal safety over raw byte optimization:

\begin{itemize}
    \item \textbf{Explicit Resource Management vs. Stack Duplication:} Standard Bitcoin Script frequently relies on \texttt{OP\_DUP} to reuse stack items (e.g., duplicating a public key for a \texttt{pk\_h} check). In contrast, Bithoven's strict \textit{liveness analysis} mandates that every resource be consumed exactly once. Consequently, operations that logically require the same value twice---such as hashing a key and then checking its signature---require distinct witness inputs (e.g., \texttt{k} and \texttt{k'}). While this increases the witness size by one stack item, it eliminates an entire class of stack manipulation vulnerabilities and ensures referential transparency.
    
    \item \textbf{Stack Hygiene:} Time-lock statements in Bithoven (\texttt{older}, \texttt{after}) compile with an explicit \texttt{OP\_DROP}. Unlike policy fragments which may leave verification results on the stack, Bithoven's imperative statements enforce a ``clean stack'' invariant. This ensures that the stack state remains predictable between sequential statements, reducing the complexity of static analysis.
\end{itemize}

In summary, where overhead exists, it is restricted primarily to the Witness data---which is discounted in Segwit and Taproot transaction weight calculations---and serves to enforce the formal safety guarantees that prevent the defect classes discussed in Section \ref{sec:evaluation}.

\subsection{Evaluation: Expressiveness and Usability} \label{sec:evaluation-expressiveness}

\begin{table*}[!t]
\centering
\caption{Comparative Analysis of Language Paradigms and Expressiveness}
\label{tab:lang-comparison}
\begin{tabular}{@{}llll@{}}
\toprule
\textbf{Feature / Property} & \textbf{Bitcoin Script} & \textbf{Miniscript} & \textbf{Bithoven} \\ \midrule
\textbf{Language Paradigm} & Stack-based VM & Composable Policy Language & Imperative, High-Level \\
\textbf{Primary Design Goal} & VM Execution & Static Analyzability & Developer Ergonomics \& Safety \\
\textbf{Intended User} & VM / Experts & Compiler / Analyzer & Human Developer \\
\textbf{Static Type System} & None & Subset-based Typing & Explicit (\texttt{num}, \texttt{string}, \texttt{sig}) \\
\textbf{Explicit State (Variables)} & None (Stack only) & None & Yes (via Stack Declaration) \\
\textbf{General Control Flow (\texttt{if/else})} & Limited (e.g., \texttt{OP\_IF}) & None (Policy-based) & Native Support \\
\textbf{Arithmetic \& Comparison} & Native (Unsafe) & Not Supported (Policy Only) & Native (Type-Safe) \\
\textbf{Arbitrary Expression Nesting} & Not Supported & Composable Fragments & Native (e.g., Math, Logic) \\
\textbf{Liveness / State Analysis} & N/A & Not Supported & Native (in Static Analyzer) \\
\textbf{Target-Specific Compilation} & N/A & N/A & Native (\texttt{pragma target}) \\ 
\bottomrule
\end{tabular}
\end{table*}

To evaluate Bithoven's expressiveness and usability, we present a case study of a standard Hashed TimeLock Contract (HTLC), comparing its implementation in raw Bitcoin Script (Listing \ref{list:htlc-script}), Miniscript (Listing \ref{list:htlc-miniscript}), and Bithoven (Listing \ref{list:htlc-bithoven}).

The raw Bitcoin Script implementation is an opaque, error-prone sequence of stack-manipulation opcodes, demonstrating the exact high-level complexity that leads to developer-introduced defects. The Miniscript policy is a monumental improvement, offering a readable, analyzable, and composable abstraction for the contract's \textit{policy}.

Bithoven bridges the gap, providing the clarity of a high-level policy with the power and intuitive structure of an imperative \textit{program}. The HTLC's two spending paths are not defined as a flat, single policy but are modeled using a familiar \texttt{if/else} control-flow block. Crucially, Bithoven's language-level design provides a feature not available in other systems: the ability to formally declare each distinct spending path with its own, separate, typed input stack. As shown in Listing \ref{list:htlc-bithoven}, the developer can reason about the ``refund'' path (taking a \texttt{sig\_alice}) and the ``redeem'' path (taking a \texttt{preimage} and \texttt{sig\_bob}) as two distinct, self-documenting, and statically-verifiable function signatures.

This case study demonstrates Bithoven's ability to express complex, multi-path logic in a way that is intuitive for developers, an advantage summarized in Table \ref{tab:lang-comparison}. Bithoven adds crucial features like imperative control flow and state-aware liveness analysis that are deliberately outside the scope of Miniscript's policy-first design.

\begin{lstlisting}[
  language={},
  caption={HTLC Contract Compiled in Bitcoin Script},
  label=list:htlc-script, captionpos=b, frame=single, float=!t,
  basicstyle=\small\ttfamily
]
OP_IF 
    OP_PUSHBYTES_N <locktime> OP_CSV OP_DROP
    OP_PUSHBYTES_33 <pk_alice> OP_CHECKSIG
OP_ELSE 
    OP_HASH256 OP_TOALTSTACK OP_PUSHBYTES_32 
    <H> OP_FROMALTSTACK OP_SWAP OP_EQUALVERIFY 
    OP_PUSHBYTES_33 <pk_bob> OP_CHECKSIG
OP_ENDIF
\end{lstlisting}

\begin{lstlisting}[
  language={},
  caption={HTLC Contract in Miniscript Policy Language},
  label=list:htlc-miniscript, captionpos=b, frame=single, float=!t,
  basicstyle=\small\ttfamily
]
or(and(pk(pk_bob), sha256(H)),
and(pk(pk_alice), older(<locktime>)))
\end{lstlisting}

\begin{lstlisting}[language=Bithoven, caption={HTLC bithoven.}, label={list:htlc-bithoven}, captionpos=b, frame=single, float=!t]
pragma bithoven version 0.0.1;
pragma bithoven target segwit;

(condition: bool, sig_alice: signature)
(condition: bool, preimage: string, sig_bob: signature)
{
    if condition {
        older <locktime>;
        return checksig (sig_alice, <pk_alice>);
    } else {
        verify sha256 sha256 preimage == <H>;
        return checksig (sig_bob, <pk_bob>);
    }
}
\end{lstlisting}


\section{Conclusion} \label{sec:conclusion}

The smart contract design space has long been defined by a stark trade-off: Ethereum's high-level expressiveness at the cost of catastrophic bugs \cite{sol-bug, byte-bug}, versus Bitcoin's robust security, which is locked behind the ``notoriously difficult'' and error-prone Bitcoin Script \cite{bitcoinorg-tx, bshunter}. This has left a gap for a pragmatic solution that is both developer-friendly and formally safe.

In this paper, we presented Bithoven, a new, formally specified, high-level imperative language for Bitcoin that fills this gap. We have demonstrated that Bithoven's core contribution is not just its intuitive syntax, but its correct-by-construction static analysis pipeline. This multi-stage analyzer---which integrates a strict type system, a novel liveness analyzer, and a semantic control-flow checker---provides \textit{a priori} safety guarantees.

Our evaluation provides three key results. First, we demonstrated that Bithoven's static analyzer prevents critical classes of documented, fund-losing vulnerabilities at compile time, including defects that Miniscript does not address, such as state-based liveness errors (e.g., duplicate signature consumption). Second, as shown in Table \ref{tab:semantics_comparison}, we dispelled the concern that high-level abstractions incur prohibitive on-chain costs. Our analysis proves that Bithoven is \textit{structurally isomorphic} to optimized Miniscript for the vast majority of operations, incurring zero logic overhead. Where minor overhead exists, it is a deliberate trade-off to enforce stack hygiene and resource safety. Finally, our comparison in Table \ref{tab:lang-comparison} highlights Bithoven's superior expressiveness, offering native support for arithmetic and imperative control flow that is absent in policy-based languages.

A central contribution of this work is Bithoven's ability to absorb logical and state-based complexity at the compiler level. By managing this complexity by construction, Bithoven proves that it is not necessary to sacrifice safety or efficiency for expressiveness. It bridges the long-standing gap between safety and usability, offering a secure, practical, and formally-grounded path for the future of Bitcoin smart contract development.

\section*{Acknowledgments}
The authors would like to acknowledge the use of Google's Gemini\cite{gemini}, a large language model, for assistance in refining the grammatical structure of the manuscript. The authors have reviewed, verified, and are fully responsible for all content, including the code and the final text presented in this article.

\bibliographystyle{IEEEtran}
\bibliography{ref}

\begin{IEEEbiography}[{\includegraphics[width=1in,height=1.25in,clip,keepaspectratio]{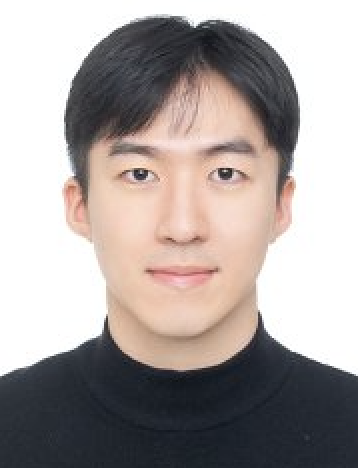}}]{Hyunhum Cho}
received the B.B.A degree in the School of Business, Yonsei University, Seoul, South Korea, in 2019, and received the M.S.
degree in industrial engineering, Seoul National University, Seoul, South Korea, in 2022. He is currently
pursuing the Ph.D. degree in information security from Korea University, Seoul, South Korea.
His main research interests include blockchain, formal methods, and cryptography.
\end{IEEEbiography}
\begin{IEEEbiography}[{\includegraphics[width=1in,height=1.25in,clip,keepaspectratio]{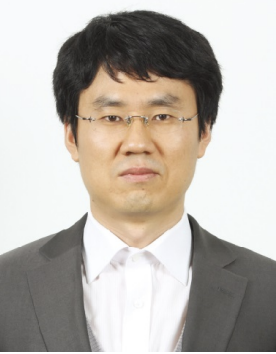}}]{Ik Rae Jeong}
received the B.S. and M.S. degrees in
computer science and the Ph.D. degree in information security from Korea University, South Korea,
in 1998, 2000, and 2004, respectively. From June
2006 to February 2008, he was a Senior Engineer
at the Electronics and Telecommunications Research
Institute (ETRI), South Korea. He is currently a
Faculty Member with the School of Cybersecurity,
Korea University. His current research interests include cryptography, theoretical computer science,
blockchain, and biometrics.
\end{IEEEbiography}

\appendices

\section{Proof of Type Safety}
\label{sec:safety-proofs}

To formally substantiate the safety claims of Bithoven, we provide the soundness proofs connecting the Type System (Section V) with the Operational Semantics (Section IV). We demonstrate that any Bithoven program accepted by the static analyzer will never reach an undefined state during execution on the Bitcoin VM.

\subsection{Definitions}

First, we define the consistency between the runtime stack and the static types.

\begin{defn}\textbf{Well-Typed Value: }
A runtime value $v$ has type $\tau$, denoted $\models v : \tau$, if:
\begin{itemize}
    \item If $\tau = \mathbf{num}$, then $v \in \mathbb{Z}$.
    \item If $\tau = \mathbf{bool}$, then $v \in \{\mathbf{true}, \mathbf{false}\}$.
    \item If $\tau = \mathbf{string}$, $\mathbf{sig}$, or $\mathbf{pubkey}$, then $v$ is a byte sequence satisfying the respective format.
\end{itemize}
\end{defn}

\begin{defn}\textbf{Well-Typed Stack: }
A stack $\sigma = v_1 :: v_2 :: \dots :: v_k$ corresponds to a type sequence $\vec{\tau} = \tau_1, \tau_2, \dots, \tau_k$, denoted $\models \sigma : \vec{\tau}$, if and only if for all $i \in \{1..k\}$, $\models v_i : \tau_i$.
\end{defn}

\subsection{Safety Theorems}

We assert two fundamental properties: \textit{Progress} (a well-typed program never gets ``stuck'') and \textit{Preservation} (execution preserves type consistency).

\begin{theorem}\textbf{Progress: }
\label{thm:progress}
Let $s$ be a statement such that $\Gamma \vdash s \Rightarrow \mathbf{ok}$. Let $\sigma$ be a stack such that $\models \sigma : \sum$ (the stack matches the input declaration). Then, the execution configuration $\langle s, \sigma \rangle$ is not stuck. It either:
\begin{enumerate}
    \item Terminates successfully with a resulting stack $\sigma'$,
    \item Terminates in a valid failure state $\perp$ (e.g., via \texttt{verify false}), or
    \item Can make a transition to an intermediate configuration $\langle s', \sigma' \rangle$.
\end{enumerate}
\end{theorem}

\begin{IEEEproof}[Proof Sketch]
We proceed by induction on the structure of the statement $s$.
\begin{itemize}
    \item \textbf{Case (Expression Evaluation):} For any expression $e$ used in $s$, the typing rules (Fig. 2) ensure operators are applied to compatible types. For instance, if the code contains $e_1 + e_2$, the rule \textsc{T-BinaryMath} ensures $e_1, e_2 : \mathbf{num}$. By the Canonical Forms lemma, the values $v_1, v_2$ must be integers. The underlying Bitcoin opcode \texttt{OP\_ADD} is defined for all integers. Thus, the operation cannot be undefined.
    \item \textbf{Case (If-Else):} For $\mathbf{if}\ e \dots$, rule \textsc{T-If} ensures $e : \mathbf{bool}$. The operational semantics define transitions for both $\mathbf{true}$ and $\mathbf{false}$. Thus, the control flow is never undefined.
    \item \textbf{Case (Locktimes):} For $\mathbf{older}\ n$, rule \textsc{T-Locktime} ensures $n$ is a valid 32-bit unsigned integer. The semantics dictate this transitions to $\sigma$ (success) or $\perp$ (failure based on transaction nSequence), both of which are valid states.
\end{itemize}
\end{IEEEproof}

\begin{theorem}\textbf{Preservation \& Subject Reduction: }
\label{thm:preservation}
If a program state is well typed, and execution takes a step, the resulting state remains well-typed. Formally, if $\Gamma \vdash s \Rightarrow \mathbf{ok}$ and $\models \sigma : \vec{\tau}_{in}$, and $\rho \vdash \langle s, \sigma \rangle \rightarrow \sigma'$, then there exists a type sequence $\vec{\tau}_{out}$ such that $\models \sigma' : \vec{\tau}_{out}$.
\end{theorem}

\begin{IEEEproof}[Proof Sketch]
We assume the induction hypothesis that preservation holds for all sub-statements.
\begin{itemize}
    \item \textbf{Base Case (Return):} $\mathbf{return}\ e$. If $\Gamma \vdash e : \tau$, then by the semantics of expressions, $e$ evaluates to a value $v$ where $\models v : \tau$. The final stack item becomes $v :: \sigma'$ (or similar depending on context), which is well-typed.
    \item \textbf{Inductive Step (Conditionals):} Consider $\mathbf{if}\ e\ \{s_1\}\ \mathbf{else}\ \{s_2\}$. The type checker enforces $\Gamma \vdash s_1 \Rightarrow \mathbf{ok}$ and $\Gamma \vdash s_2 \Rightarrow \mathbf{ok}$. If $e$ evaluates to $\mathbf{true}$, we step into $s_1$. By the inductive hypothesis, since $s_1$ is well-typed, its execution results in a well-typed stack $\sigma'$. The same applies if $e$ is $\mathbf{false}$ for $s_2$.
    \item \textbf{Inductive Step (Verify):} $\mathbf{verify}\ e$. The semantics state that if $e \to \mathbf{true}$, the stack remains $\sigma'$ (the state after popping $e$). Since the stack was well-typed before the push/pop of the boolean, it remains well-typed. If $e \to \mathbf{false}$, the state becomes $\perp$, which is a valid terminal state.
\end{itemize}
Consequently, Bithoven programs do not exhibit type confusion errors during execution.
\end{IEEEproof}

\section{Catalogue of Prevented Defects}
This appendix provides a comprehensive catalogue of defect classes that Bithoven's static analyzer prevents at compile time. Each example corresponds to a vulnerability class identified in Table II, demonstrating the analyzer's robustness.

\subsection{Vulnerability Class: Type-System Safety}

\subsubsection{Defect: Arithmetic on Non-Numeric Types}

The developer attempts to apply mathematical operators (\texttt{+} and \texttt{-}) to string literals, which is disallowed in Bitcoin Script.
\begin{lstlisting}[language=Bithoven, caption=Applying math operators to strings., label=list:app-math-on-string]
pragma bithoven version 0.0.1;
pragma bithoven target segwit;

(sig_alice: signature)
{
verify ("MATH" + "ONLY FOR" - "NUMERIC");
}
\end{lstlisting}

The type checker correctly identifies that the \texttt{+} and \texttt{-} operations on \texttt{string}, which is the wrong type for arithmetic operation.
\compilererror{
Error at line 6:13: InvalidOperation("Operand must be number or boolean but: StringLiteral(Location { start: 100, end: 106, line: 6, column: 13 }, \string"MATH\string")")
}

\textbf{   }
\subsubsection{Defect: Comparison of Mismatched Types}
The developer attempts to compare a number literal \texttt{2} with a string literal.
\begin{lstlisting}[language=Bithoven, caption=Comparing a number to a string., label=list:app-compare-mismatch]
pragma bithoven version 0.0.1;
pragma bithoven target segwit;

(sig_alice: signature)
{
return (2 < "Wrong Type");
}
\end{lstlisting}

The type checker rejects this operation, enforcing that comparisons can only occur between values of the same type.
\compilererror{
Error at line 6:13: InvalidOperation("Compare type must be same but: NumberLiteral(Location { start: 100, end: 101, line: 6, column: 13 }, 2) to StringLiteral(Location { start: 104, end: 116, line: 6, column: 17 }, \string"Wrong Type\string")")
}

\subsection{Vulnerability Class: Liveness \& Scope}

\subsubsection{Defect: Referencing an Undefined Variable}
The stack declaration defines \texttt{sig\_alice}, but the code body mistakenly references \texttt{sig\_bob}.
\begin{lstlisting}[language=Bithoven, caption=Using an undefined variable., label=list:app-undefined-var]
pragma bithoven version 0.0.1;
pragma bithoven target segwit;

(sig_alice: signature)
{
return checksig (sig_bob, "0245...bf5212");
}
\end{lstlisting}

The liveness and scope analyzer immediately flags that \texttt{sig\_bob} has not been declared in the current scope.
\compilererror{
Error at line 6:22: UndefinedVariable("Undefined variable: \string"sig\_bob\string".")
}

\subsection{Vulnerability Class: Semantic Security}

\subsubsection{Defect: Mangled Logic (Useless Signature Check)}
This corresponds to the \texttt{useless-sig} vulnerability class from \cite{bshunter}.
The developer inverts the result of \texttt{checksig}. This is a critical flaw, as an adversary can now spend the funds by providing an invalid signature, which causes \texttt{checksig} to return \texttt{false}, which is then inverted to \texttt{true} by the \texttt{!} operator.
\begin{lstlisting}[language=Bithoven, caption=Inverting a checksig result., label=list:app-useless-sig]
pragma bithoven version 0.0.1;
pragma bithoven target segwit;

(sig_alice: signature)
{
return ! checksig (sig_alice, "0245...bf5212");
}
\end{lstlisting}

Bithoven's semantic security analyzer is designed to find this exact anti-pattern, preventing mangled cryptographic logic by construction.
\compilererror{
Error at line 6:12: UselessSig("! makes checksig operation useless: UnaryMathExpression { loc: Location { start: 99, end: 198, line: 6, column: 12 }, operand: CheckSigExpression { loc: Location { start: 101, end: 198, line: 6, column: 14 }, operand: SingleSigFactor { loc: Location { start: 110, end: 198, line: 6, column: 23 }, sig: Variable(Location { start: 111, end: 120, line: 6, column: 24 }, Identifier(\string"sig\_alice\string")), pubkey: StringLiteral(Location { start: 123, end: 197, line: 7, column: 7 }, \string"0245a...f5212\string") }, op: CheckSig }, op: Not }.")
}

\text{  }

\subsubsection{Defect: Uncertain Signature (uncertain-sig)}
The developer attempts to validate a signature using a public key provided dynamically on the stack (\texttt{pubkey\_alice}), rather than a static constant. This allows an attacker to supply their own key to satisfy the check.

\begin{lstlisting}[language=Bithoven, caption=Attempting to use a dynamic public key., label=lst:app-uncertain-sig]
pragma bithoven version 0.0.1;
pragma bithoven target segwit;

(sig_alice: signature, pubkey_alice: string)
{
  return checksig(sig_alice, pubkey_alice);
}
\end{lstlisting}

The type checker rejects this operation, enforcing that public keys must be static literals to prevent key substitution attacks.

\compilererror{
Error at line 6:33: TypeMismatch("Public Key must be from string literal but: Variable(Location \{ start: 142, end: 154, line: 6, column: 33 \}, Identifier(\string"pubkey\_alice\string")).")
}

\subsection{Vulnerability Class: Consensus Rules}

\subsubsection{Defect: Integer Overflow (32-bit Limit)}
Bitcoin Script have implicit 32-bit integer limits.
The developer provides a number literal (\texttt{21474836478}) that exceeds the maximum value of a 32-bit signed magnitude integer (\texttt{2147483647}).
\begin{lstlisting}[language=Bithoven, caption=Integer literal out of range., label=list:app-int-overflow]
pragma bithoven version 0.0.1;
pragma bithoven target segwit;

(sig_alice: signature)
{
return 21474836478;
}
\end{lstlisting}

The compiler's literal validation prevents this consensus-level defect before deployment.
\compilererror{
Error at line 6:12: IntegerOverflow("Number is 32 bit sign magnitude int: 21474836478")
}

\subsection{Vulnerability Class: Control Flow}

\subsubsection{Defect: Missing Return Statement}
This corresponds to the \texttt{never-true} defect class from \cite{bshunter}, as the script can finish without returning a \texttt{true} value.
The code path provides \texttt{verify} and \texttt{older} statements but is missing a terminal \texttt{return} statement. The script would execute and then fail, as the stack would be empty at termination.
\begin{lstlisting}[language=Bithoven, caption=Missing a terminal return statement., label=list:app-no-return]
pragma bithoven version 0.0.1;
pragma bithoven target segwit;

(sig_alice: signature)
{
verify checksig(sig_alice, "0245...bf5212");
older 1000;
}
\end{lstlisting}

The control-flow analyzer traverses the program's CFG and detects that a valid path exists which does not terminate in a \texttt{return}.
\texttt{ }
\compilererror{
Error at line 7:5: NoReturn("Return statement must exist for each possible execution path: LocktimeStatement { loc: Location { start: 195, end: 211, line: 7, column: 5 }, operand: 1000, op: Csv }.")
}
\texttt{ }

\subsubsection{Defect: Unreachable Code}

The developer has placed an \texttt{older} statement after the terminal \texttt{return} statement, making it impossible to execute.
\begin{lstlisting}[language=Bithoven, caption=Code placed after a return statement., label=list:app-unreachable]
pragma bithoven version 0.0.1;
pragma bithoven target segwit;

(sig_alice: signature)
{
return checksig (sig_alice, "0245...bf5212");
older 1000;
}
\end{lstlisting}

The control-flow analyzer identifies all code that follows a terminal operation in the same block as unreachable.
\compilererror{
Error at line 7:5: UnreachableCode("Unreachable code after return statement: LocktimeStatement { loc: Location { start: 195, end: 211, line: 7, column: 5 }, operand: 1000, op: Csv }. Move return statement at the last scope of execution path")
}

\end{document}